\documentclass[pdflatex,sn-mathphys-num]{sn-jnl}% Math and Physical Sciences Numbered Reference Style
\usepackage{graphicx}%
\usepackage{multirow}%
\usepackage{amsmath,amssymb,amsfonts}%
\usepackage{amsthm}%
\usepackage{mathrsfs}%
\usepackage[title]{appendix}%
\usepackage{xcolor}%
\usepackage{textcomp}%
\usepackage{manyfoot}%
\usepackage{booktabs}%
\usepackage{algorithm}%
\usepackage{algorithmicx}%
\usepackage{algpseudocode}%
\usepackage{listings}%
\usepackage[numbers]{natbib}
\usepackage[utf8]{inputenc}
\theoremstyle{thmstyleone}%
\theoremstyle{thmstyletwo}%

\theoremstyle{thmstylethree}%

\begin{document}

\title[Article Title]{Semiconductor--Semimetal Transition in van der Waals Carbyne Crystals}


\author*[1,2]{\fnm{Daniele} \sur{Barettin}}\email{daniele.barettin@unicusano.it}

\author*[2,3]{\fnm{Stella V.} \sur{Kavokina}}\email{kutrovskaya.sv@mipt.ru}
\author[4]{\fnm{Evgeny L.} \sur{Ivchenko}}\email{ivchenko@coherent.ioffe.ru}

\author[2,3]{\fnm{Alexey V.} \sur{Kavokin}}\email{kavokin.av@mipt.ru}

\affil[1]{\orgdiv{Department of Electronic Engineering}, \orgname{Universit\`a Niccol\`o Cusano},\\ \orgaddress{\street{Via Don Carlo Gnocchi}, \city{Rome}, \postcode{00166},\country{Italy}}}

\affil[2]{\orgdiv{Abrikosov Center for Theoretical Physics}, \orgname{Moscow Institute of Physics and Technology}, \orgaddress{\street{\mbox{}}\city{Dolgoprudnyi}, \postcode{141701},  \country{Russia}}}

\affil[3]{\orgdiv{\mbox{}}\orgname{Saint-Petersburg State University Institute},\orgaddress{\street{\mbox{}} \city{St.~Petersburg}, \postcode{198504}, \country{Russia}}}

\affil[4]{\orgdiv{\mbox{}}\orgname{Ioffe Institute},\orgaddress{\street{\mbox{}} \city{St.~Petersburg}, \postcode{194021}, \country{Russia}}}

\abstract{Freestanding van der Waals crystals made of single-atom carbon chains (carbynes) have been recently realized technologically. Here we investigate their electronic and optical properties experimentally, by continuous-wave and time-resolved photoluminescence spectroscopy, and theoretically. Employing a fully three-dimensional tight-binding formalism benchmarked against density functional theory calculations we predict the semimetal-semiconductor transition to occur in van der Waals carbyne crystals composed by the chains of about 42 atoms long. The semiconductor phase is characterized by a hyperbolic van Hove singularity which gives rise to unconventional hyperbolic exciton states. Experimentally, we access the semiconductor phase, where resonant features associated with hyperbolic excitons are clearly visible. The exciton oscillator strength is found to be strongly sensitive to the length of carbon chains: our experiments show that it decreases with the increasing chain length. This tendency, confirmed by the theoretical modeling, manifests the evolution of the hyperbolic exciton state on the way to the semiconductor-semimetal crossover. Our approach accounts for the actual crystalline geometry, including alternating intra-chain hoppings and inter-chain couplings. By fitting tight-binding dispersions to density functional theory data we extract consistent parameters and establish a comprehensive framework for the physics of carbyne crystals. This study paves the way towards efficient band-gap engineering in ultimate one-dimensional carbon crystals.}




\maketitle

\section{Introduction}\label{sec1}

The crystal phase transition between a semiconductor band-structure characterized by a positive energy gap and a semi-metal band structure characterized by overlapping conduction and valence bands and negative effective band-gap are in the focus of intense studies since 1960s ~\cite{Halperin1968}. A dramatic change of electronic and optical properties of crystals undergoing such transitions as well as fascinating topological effects in the vicinity of a transition point make semimetal-semiconductor transitions (SST) interesting from the fundamental point of view and important for applications ~\cite{TeSn2017, TMD2019}. The SST may be controlled by pressure~\cite{Halperin1968}, electrostatic gating~\cite{Reed2016}, temperature~\cite{MoTe2025} or chemical compound of an alloy~\cite{Prakash2025}. Being characterized by specific sets of critical parameters and well defined transition points, SSTs need to be distinguished from the gap opening phenomena triggered by size quantization in finite size crystals, such as graphene nanoribbons~\cite{Nanoribbon2021}. Indeed, formally, any nanoribbon would possess an energy gap, there is no phase transition between a gapless graphene and gapped nano-ribbon band structures. 

In this work, we study theoretically and experimentally SST that is controlled by a size of a crystal but can be characterized by a specific transition point. This transition appears to be one of unique features one Van der Waals crystals composed on one-dimensional sp-carbon chains or carbynes.

Carbyne, the ultimate one-dimensional carbon allotrope, has long fascinated theorists due to its anticipated unique electronic and structural properties but remained elusive because of its instability in bulk form ~\cite{Kasatochkin1967, Goresy1968, Whittaker1969}. 
Carbyne can exist in two structural isomers: cumulene, with equal bond lengths and metallic character, and polyyne, in which alternating single and triple bonds arise from a Peierls distortion, opening a finite gap~\cite{Heimann1999, Chalifoux2010}. 
Recent progress in chemical stabilization strategies - including encapsulation in carbon nanotubes, polymer-derived synthesis, and anchoring to gold nanoparticles - has made it possible to stabilize finite carbyne chains, reviving intense interest in their electronic and optical properties~\cite{Tykwinski, Casari2016, Kutrovskaya2020}. 

Experimental studies have now provided unambiguous evidence of semiconducting behavior of short polyyne chains. 
Indeed, the polyyne chains stabilized by gold nanoparticles display strong visible photoluminescence (PL), with emission energies tunable between 2 and 4~eV depending on the chain length~\cite{Kutrovskaya2020}. 
At cryogenic temperatures, the PL spectra show pronounced fine structure, including sharp excitonic peaks and trion sidebands, while time-resolved PL reveals radiative lifetimes on the order of 1~ns that increase systematically with chain length. The effective band gap is governed by the length of straight chain segments, confined between kinks or between a kink and the metallic surface, which always contain an even number of carbon atoms. 
For segments with 8-24 atoms, the measured band gaps fall in the 1.93-2.95~eV range, in excellent agreement with the presence of strongly bound excitons~\cite{Kutrovskaya2021}. 
Further investigations including the X-ray and electron diffraction spectroscopy clarified that the chains of sp-carbon atoms form haxagonal van der Waals crystals characterized by a lattice constant of about 0.535 nm (to be compared with much smaller interatomic distances within the individual chains, estimated to be 0.123 and 0.133 nm for the carbon atoms separated by triple and single electronic bonds, respectively ~\cite{Kutrovskaya2020}. 
The coupling of parallel carbyne chains was studied theoretically by Ara and Basu~\cite{coupled}, however, there is no published theoretical studies on the band structure, electronic and optical properties of van der Waals carbyne crystals known to us. To fill this gap, we have performed a careful theoretical study of hexagonal carbyne crystals and confronted its results to the original experimental data extracted from low temperature cw and time-resolved photoluminescence spectra of carbyne crystals. This study led us to a surprising conclusion: in contrast to single polyyne chains, van der Waals crystals made of polyyne chains are expected to exhibit a semiconductor-semimetal phase transition (SST). The transition is governed by the lengths of the chains composing the crystal. According to our calculations, the carbyne crystal is a direct gap semiconductor if composed by the polyyne chains shorter than 42 atoms, and it represents a semimetal characterized by a zero energy gap otherwise.

Experimentally, we had access to the semiconductor samples composed by the chains of 8 to 24 atoms. They feature strong exciton-resonances whose energy and oscillator strength rapidly decrease with the increase of the length of the chain. Both tendencies appear to be in excellent agreement with our theoretical results. Noteworthy, carbyne crystals are characterized by unusual, hyperbolic excitons formed by electrons and holes located in van Hove singular points of the conduction and valence bands, respectively. The experimental study of the optical properties of the hyperbolic excitons in carbyne crystals confirms the anticipated tendency to the semiconductor-semimetal crossover. These results represent the first direct observation of bright excitons in monoatomic carbon chains and emphasize the decisive role of finite-size quantization in their optical response. The control of the length of the chain appears to be an excellent tool for band engineering in ultimate one-dimensional carbon crystals.

On the theoretical side, one-dimensional (1D) tight-binding (TB) models have long served as a minimal description of carbyne, capturing the Peierls instability and band-gap opening~\cite{polyyne}. 
Recently, Ara and Basu~\cite{coupled} extended the TB analysis to coupled polyyne chains, showing that inter-chain interactions may drive topological phases in 1D arrays. 
However, such approaches neglect the genuine three-dimensional (3D) geometry of van der Waals carbyne crystals, where parallel chains form hexagonal arrays in the $xy$ plane and extend along the $z$ axis. 
In this setting, inter-chain tunneling plays a critical role in determining the electronic dispersion and gap structure, and cannot be reduced to a purely 1D description.

Two distinct stacking arrangements are possible for van der Waals carbyne crystals (see Figure 1): in the AA configuration, adjacent chains are aligned without relative shift along the $z$ axis, whereas in the AB configuration, neighboring chains are displaced by half a period, leading to a different registry of the alternating single and triple bonds. Experimentally, both stackings are likely to be present in the available samples, having in mind that kinks occurring at different locations of the parallel chains help switching from AA to AB stacking and backward. 
In this work, we develop a fully 3D TB model for van der Waals carbyne crystals, benchmarked against ab initio density functional theory (DFT) simulations for both AA and AB stacking geometries. 
This combined approach enables us to extract consistent hopping parameters and to capture the interplay between intra-chain Peierls distortion and inter-chain coupling. 

Our results demonstrate the SST driven by chain length: finite chains behave as semiconductors, but beyond $N\sim 42$ atoms the band gap closes, giving rise to a semimetallic state. 
This transition, absent in single polyyne chains, is further supported by comparison with PL data. 
In addition, we identify a hyperbolic van Hove singularity in the AA configuration, leading to unconventional hyperbolic exciton states~\cite{Velicky1966}. 
Altogether, our combined experimental, TB, and DFT analysis provides a comprehensive framework for the physics of van der Waals carbyne crystals and establishes length-controlled band-gap engineering as a central paradigm for this unique carbon allotrope.

\section{Experimental Methods}

We synthesized polyyne chains stabilized by gold nanoparticles attached to their ends with use of the laser ablation in liquid technique as described in Ref.~\cite{Kutrovskaya2020}. Both AA and AB stackings of the polyyne chain are likely to be present in our system. Indeed, close to the gold surface AA stacking must prevail, as single electronic bonds of all chains are expected to be attached to the gold surface. However, at a distance of several nanometers from the gold surface the stacking may be changed because of kinks on the chains. Due to the kinks (displacements of carbon atoms), the carbon thread of about 100-500 nm length connecting different gold nanoparticles is divided into several tens of fragments, each composed by straight carbon chains. The type of stacking is expected to change between fragments stochastically. For optical studies, the chains have been deposited on a silica glass substrate by sputtering in the presence of external electric field. Time- and spectrally-resolved PL spectra have been acquired at the liquid helium temperature under laser excitation at the wavelength of 390 nm. For details of the spectral analysis we refer the reader to our previous publication~\cite{Kutrovskaya2021}.

\section{Theoretical Models}
\subsection{Three-Dimensional Tight-Binding Approach}
\label{TModels}

To capture the essential physics of van der Waals carbyne crystals, we adopt a fully three-dimensional tight-binding (3D TB) description. Unlike earlier one-dimensional (1D) approaches~\cite{coupled,polyyne}, this formalism explicitly accounts for the crystalline geometry: linear carbon chains arranged on a hexagonal lattice in the $xy$ plane and stacked along the $z$ axis. The model naturally incorporates both alternating intra-chain hoppings $(t_1,t_2)$ and inter-chain couplings $\gamma$, providing an analytical counterpart to our DFT calculations.

\begin{figure}[ht]
    \centering
    \includegraphics[width=0.9\textwidth]{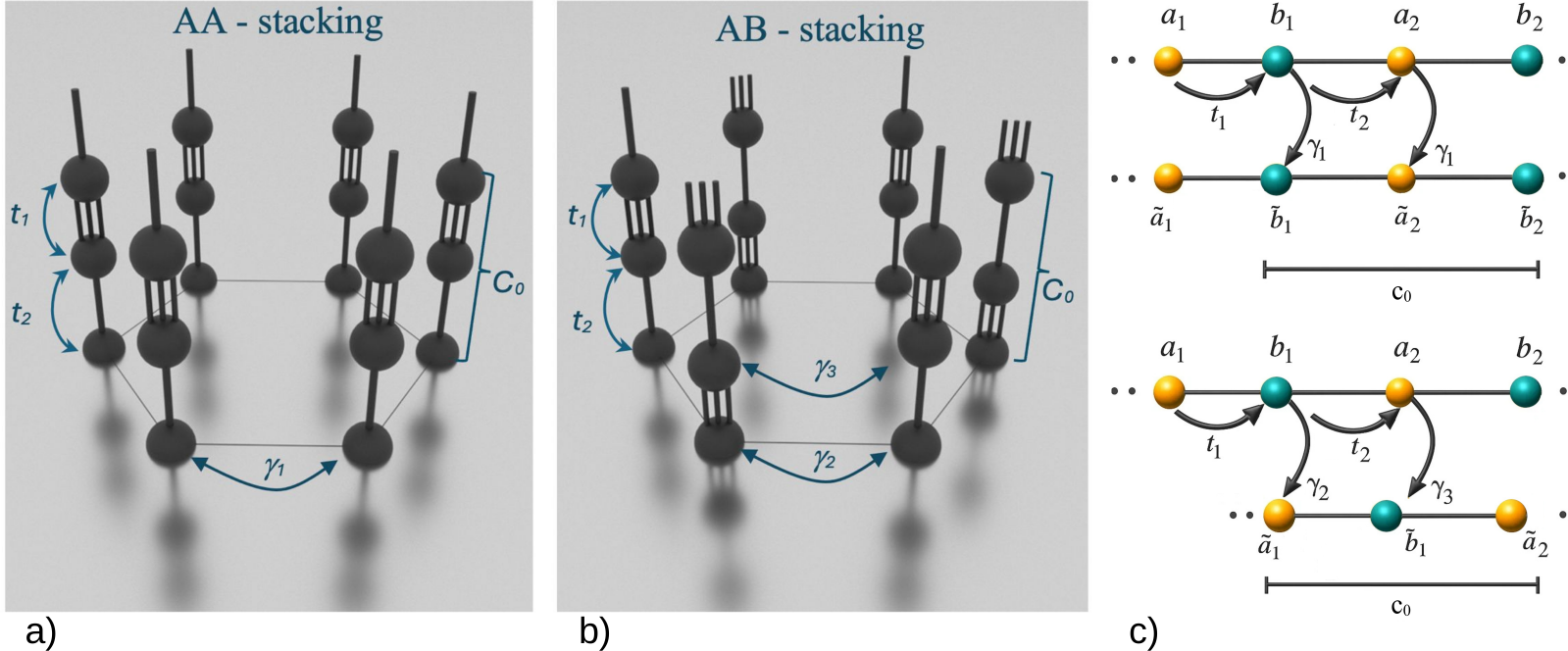}
    \caption{Schematic representation of the van der Waals carbyne crystal in the two stacking geometries. 
    a) Three-dimensional view of the AA stacking, with intra-chain hoppings $t_1$, $t_2$ and inter-chain coupling $\gamma_1$. 
    b) Three-dimensional view of the AB stacking, where the nonequivalent couplings $\gamma_2$ and $\gamma_3$ are included. 
    c) Two-dimensional schematic of bond alternation, with atoms $a_i,b_i$ (first chain) and $\tilde{a}_i,\tilde{b}_i$ (second chain) labeled explicitly.}
    \label{fig:models}
\end{figure}

For the AA stacking, the $4\times4$ TB Hamiltonian reads
\begin{equation}
\mathcal{H}(\mathbf{k})=
\begin{bmatrix}
E_0 & h_z(k_z) & h_{xy}(\mathbf{k}_\parallel) & 0\\
h_z^*(k_z) & E_0 & 0 & h_{xy}(\mathbf{k}_\parallel)\\
h_{xy}^*(\mathbf{k}_\parallel) & 0 & E_0 & h_z(k_z)\\
0 & h_{xy}^*(\mathbf{k}_\parallel) & h_z^*(k_z) & E_0
\end{bmatrix},
\label{eq:H-4x4-AA}
\end{equation}
with
\begin{equation}
h_z(k_z)= t_1 e^{i k_z c_0/2}+ t_2 e^{-i k_z c_0/2}, \qquad
h_{xy}(\mathbf{k}_\parallel)= \gamma_1 \sum_{n=1}^{3} e^{i\mathbf{k}_\parallel\cdot \mathbf{r}_n}\:,
\label{eq:hz-hxy-def}
\end{equation}
where $\mathbf{k}_\parallel$ is the in-plane component of the electron wave vector $\mathbf{k}$, and $\mathbf{r}_n$ are the positions of three nearest in-plane neighbors.

For the AB stacking, where neighboring chains are shifted by half a period along $z$, the Hamiltonian becomes
\begin{equation}
\mathcal{H}(\mathbf{k})=
\begin{bmatrix}
E_0 & h_z(k_z) & 0 & e^{-i k_z c_0}\,\gamma_2\,\eta_{xy}(\mathbf{k}_\parallel)\\
h_z^*(k_z) & E_0 & \gamma_3\,\eta_{xy}(\mathbf{k}_\parallel) & 0\\
0 & \gamma_3\,\eta_{xy}^*(\mathbf{k}_\parallel) & E_0 & h_z(k_z)\\
e^{i k_z c_0}\,\gamma_2\,\eta_{xy}^*(\mathbf{k}_\parallel) & 0 & h_z^*(k_z) & E_0
\end{bmatrix},
\label{eq:H-4x4-AB}
\end{equation}
where $\eta_{xy}(\mathbf{k}_\parallel)=\sum_{n=1}^3 e^{i\mathbf{k}_\parallel\cdot \mathbf{r}_n}$.

The full derivation of the Hamiltonians and dispersion relations are provided in the Supplementary Materials.

\subsection{Density Functional Theory: Computational Details}

The TB parameters for both AA and AB stackings were extracted from fully 3D DFT calculations performed using the PBE functional as implemented in the \textsc{Quantum~ESPRESSO} package~\cite{QE2009, QE2017}. 
In contrast to previous 1D models, the present simulations employed the actual 3D crystalline geometry: the first atoms of each chain are arranged in a hexagonal structure, from which individual linear chains extend from each vertex, reproducing the crystal structure used also in our analytical TB model. 

For the AA stacking, the simulation cell was defined with lattice parameters $a_0 = 6.15$~\AA\ and $c_0 = 2.56$~\AA~\cite{coupled}, containing four carbon atoms per cell. 
Brillouin-zone sampling along the chain axis employed a path discretized with 30 $k$-points. A plane-wave cutoff of 50~Ry and a Gaussian smearing of 0.01~Ry to facilitate convergence were used.
The AB stacking was modeled analogously, with alternating single-triple bonds shifted along the chain direction. 
The TB hopping parameters $t_1$, $t_2$, and $\gamma$ were subsequently obtained by least-squares fitting of the analytical TB dispersion to the DFT-computed bands.

\section{Results and Discussion}
\subsection{Electronic Band Structure}
\label{Results1}
\begin{figure}[ht]
    \centering
    \includegraphics[width=0.95\textwidth]{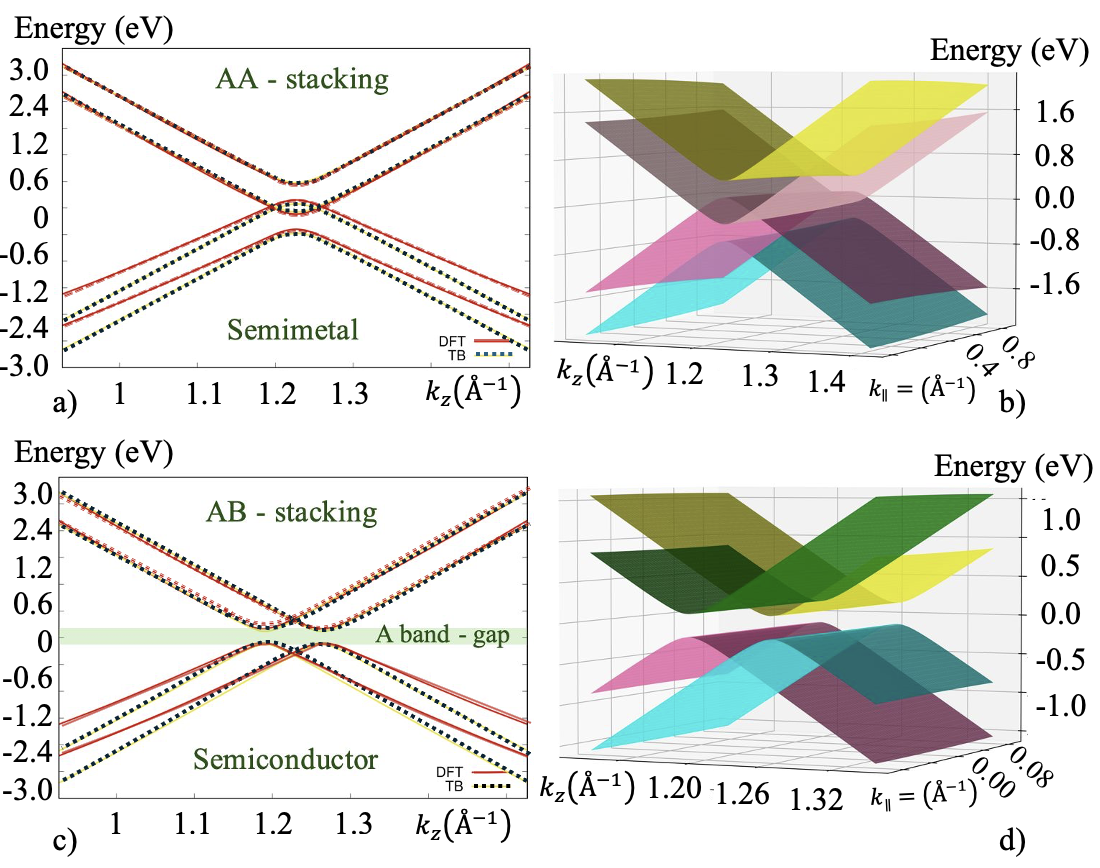}
\caption{Electronic band structures of van der Waals carbyne crystals.
(a) Comparison of DFT and TB band dispersions along the $k_z$ direction for the AA stacking.
(b) Three-dimensional representation of the electronic band structure for the AA stacking obtained from the tight-binding model.
(c) Comparison of DFT and TB band dispersions along the $k_z$ direction for the AB stacking.
(d) Three-dimensional representation of the electronic band structure for the AB stacking obtained from the tight-binding model.
The fitted TB parameters are summarized in Table~\ref{tab:TBparams}.}
    \label{fig:CompAA}
\end{figure}

In this section we focus on the electronic band structure of infinite carbon chains arranged in van der Waals crystals. 
The analysis is carried out for the two possible stacking geometries, AA and AB. We first address the electronic structure
of the AA stacking configuration. 
Figure~\ref{fig:CompAA} a) compares the DFT-calculated dispersion along the $k_z$ direction with the corresponding TB results. 
The agreement is very good, confirming that a three-parameter TB model accurately reproduces the band structure obtained from first-principles.

\begin{table}[h]
   \caption{Tight-binding parameters (in eV) extracted from the fit to DFT band dispersions.}   \label{tab:TBparams}
    \begin{tabular} {@{}lllll@{}}
        \toprule
        Stacking & $t_1$ & $t_2$ & $\gamma_1$ & $\gamma_2$\\
        AA & 3.51954 & 3.70851 & 0.104& --- \\
        AB &  3.78 & 3.92 & --- & 0.12 \\ % placeholder for later
        \botrule
    \end{tabular}
\end{table}

From the fit we obtain the TB parameters reported in Table~\ref{tab:TBparams}, 
and in Figure~\ref{fig:CompAA} b) we present the three-dimensional band dispersion for the AA configuration, which clearly illustrates the separation between the longitudinal $k_z$ dependence and the in-plane $k_\parallel$ contribution we previously discussed.

As evident from both Figure~\ref{fig:CompAA} a) and the three-dimensional representation in Figure~\ref{fig:CompAA} b), the most prominent effect of the inter-chain coupling, governed by the parameter $\gamma_1$, is to lower the energy of the lowest conduction band and simultaneously raise that of the highest valence band. 
This interaction effectively closes the band gap, driving the infinite-chain AA structure towards a metallic behavior.

We next examine the AB stacking configuration, whose DFT band structure is found to be quantitatively reproduced by the tight-binding model with equal inter-chain couplings ($\gamma_2=\gamma_3$). 
The comparison between DFT and TB dispersions along the $k_z$ direction, displayed in Figure~\ref{fig:CompAA} c), confirms the excellent agreement obtained with this parametrization. 
The corresponding fitted parameters are reported in Table~\ref{tab:TBparams} as well. 

The 3D band dispersion from the TB model is shown in Fig.~\ref{fig:CompAA} d).
In this representation the shift of the bands induced by the relative displacement of the chains becomes clearly visible. 
In contrast to the AA configuration, the effect of the inter-chain coupling $\gamma$ in the AB case is not to lower or raise the band-edge energies, but rather to produce a symmetric shift along $k_z$ of the two lowest conduction bands and the two highest valence bands.

It is worth noting that the more general case with $\gamma_2 \neq \gamma_3$ was found not to be physically relevant for the present system and is therefore not discussed further. 
%Moreover, experimental evidence indicates that the AB stacking is far less common than the AA configuration, which represents the majority of finite chains obtained in experimental growth.
%Consequently, in the following we shall focus on the most relevant features of the AA case only.

Having established the accuracy of the 3D TB model and its consistency with DFT for both stacking configurations, we now turn to one of the central results of this work: the semiconductor-semimetal transition that emerges in the AA case as the chain length increases. 
This transition, rooted in the progressive closing of the band gap induced by inter-chain coupling, marks the crossover from a Peierls-distorted semiconductor to a metallic regime in the limit of infinite chains.

\section{Semiconductor-to-Semimetal Transition in Finite Chains}

Experimental studies on polyyne chains stabilized by gold nanoparticles have provided strong evidence that these systems behave as direct-gap semiconductors with strong visible PL. 
Room-temperature spectra show broad resonances whose positions shift to lower energies as the length of the chain increases, consistent with the expected gap scaling from 2 to 4~eV for chains containing 8-22 atoms~\cite{Kutrovskaya2020}. 
At cryogenic temperatures, each PL band develops a remarkable fine structure composed of a sharp exciton peak flanked by two satellite lines that can be attributed to negatively and positively charged trions. Time-resolved measurements further reveal radiative lifetimes on the order of 1~ns, comparable to those observed in carbon nanotubes, and increasing systematically with chain length. 

Subsequent experiments have clarified that the effective band gap is determined by the length of the straight segments of the chain confined between two kinks or between a kink and the gold surface. 
Such segments invariably contain an even number of carbon atoms, and their band gap varies between 2.2 and 2.8~eV for 10-20 atom chains~\cite{Kutrovskaya2021}. 
A simultaneous decrease of the radiative lifetime with an increasing gap has been reported, in agreement with nonlocal dielectric response theory. 
These results represent the first clear observation of bright exciton states in monoatomic carbon chains and demonstrate that finite-size effects play a decisive role in shaping their optical response. 

%Overall, the data strongly suggest that AA stacking is the dominant structural motif, consistent with chains anchored to Au via single bonds. They also point to a semiconductor-semimetal transition that may occur only beyond a critical chain length, estimated experimentally in the range of 32-36 atoms. 

An SST transition in van der Waals carbyne crystals cannot be captured by a simplified one-dimensional model and requires the full 3D TB formalism introduced above.
According to Eq.~(\ref{eq:H-4x4-AA}), within this framework the dispersion for AA stacking takes the form
\begin{equation}
E_{\alpha,\beta}(\mathbf{k}) = E_0 + \alpha |h_z(k_z)| + \beta |h_{xy}(\mathbf{k}_\parallel)|,
\qquad \alpha,\beta = \pm,
\label{eq:Eaab}
\end{equation}
where 
\begin{eqnarray} \label{eq:hzhxy}
&&|h_z(k_z)| = \sqrt{t_1^2+t_2^2+2t_1t_2\cos(k_z c_0)}, \\
&& |h_{xy}(\mathbf{k}_\parallel)| = \gamma_1 \!\left\vert e^{-i k_y a_0/\sqrt{3}} + 2e^{i k_y a_0/(2\sqrt{3})}\cos{(k_x a_0/2)}\right\vert. \nonumber
\end{eqnarray}

Along the $k_z$ direction the one-dimensional dispersion features a gap of width $2|t_1-t_2|$. 
Since $|h_{xy}(\mathbf{k}_\perp)|$ spans the interval $(0,3\gamma_1)$, the condition for the gap to remain open is that $3\gamma_1 < |t_1-t_2|$. 
If this inequality is violated, the dispersion in the transverse direction closes the gap, leading to a metallic spectrum. 
The prefactor 3 originates from the maximum of $|h_{xy}|$ at $\mathbf{k}_\parallel=0$.

\vspace{0.3cm}
For finite chains, quantization along $z$ modifies the gap energy as follow: 
\begin{equation}
E_{\mathrm{gap}}(N) = 2 \sqrt{(t_1 - t_2)^2 + 4 t_1 t_2 \sin^2\!\left(\tfrac{\pi}{2N}\right)} - 6 \gamma_1 ,
\label{eq:Eg_finite}
\end{equation}
where $N$ is the number of atoms per chain. 
A positive value of Eq.~\eqref{eq:Eg_finite} corresponds to a semiconducting regime, while a negative value indicates a gap closure and the onset of semimetallic behavior. 

\begin{figure}[ht]
    \centering
    \includegraphics[width=0.9\linewidth]{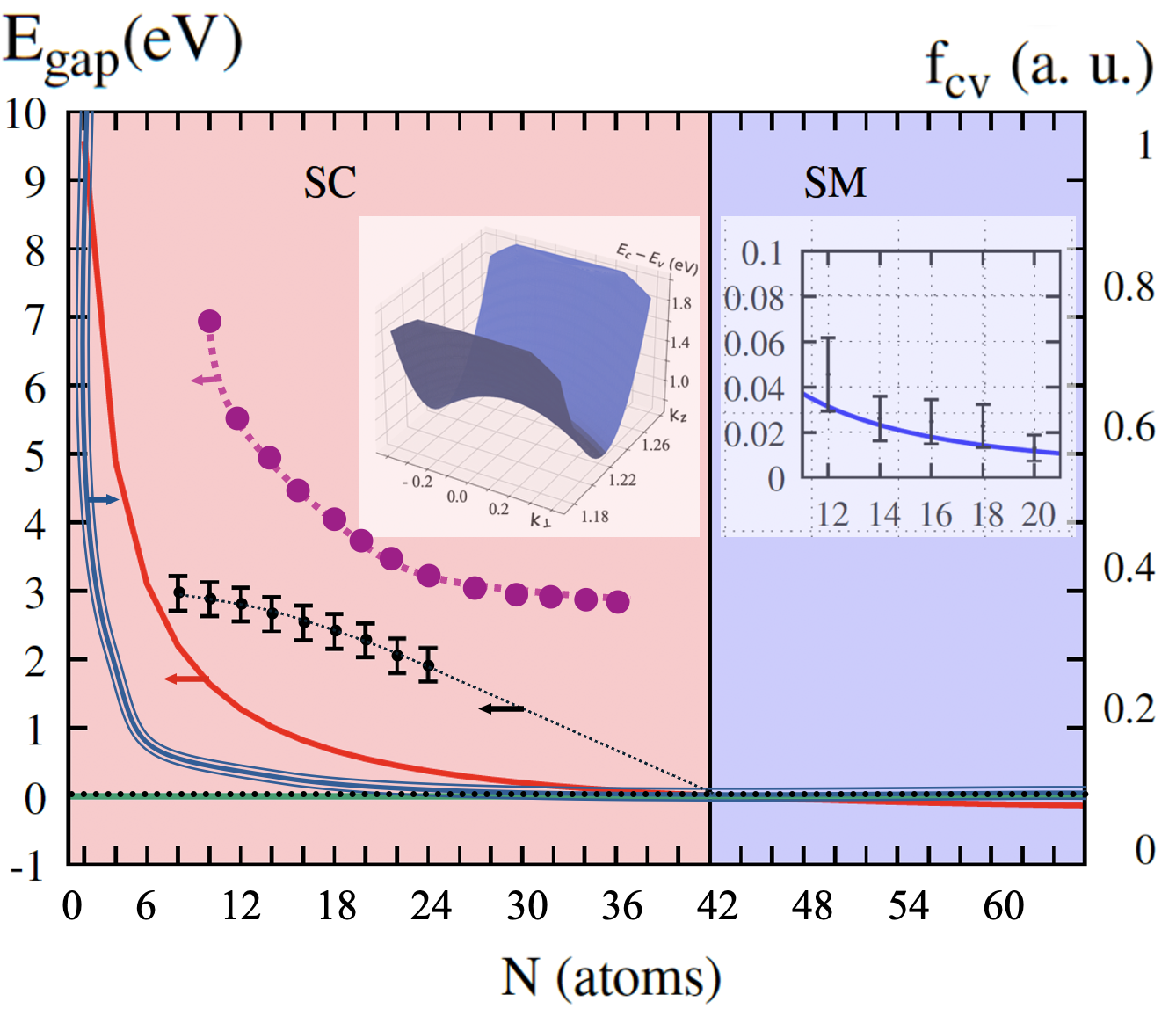}
  \caption{ Finite-size band gap $E_{\mathrm{gap}}(N)$ as a function of the chain length $N$, obtained from Eq.~\eqref{eq:Eg_finite} (red curve, left scale). 
The change of sign of $E_{\mathrm{gap}}$ indicates the semiconductor-to-semimetal transition, which occurs at $N \approx 42$. 
Points show the experimental PL data from Refs.~\cite{Kutrovskaya2020,Kutrovskaya2021} with black curve serving as a guide for the eye. The violet curve and points show the predictions of a tight-binding model for a single finite polyyne chain confined with gold atoms. The data are taken from Refs.~\cite{Portnoi}. The blue curve (right scale) shows the oscillator strength $f_{cv}(N)$ of the hyperbolic exciton plotted as a function of chain length $N$ (normalized units). The left inset shows the saddle point $(0,0,\pi/c_0)$ of the conduction band associated with the considered exciton resonance. The right inset shows the zoom on theoretical dependence of the exciton oscillator strength on the length of the chains compared to the experimental data obtained by time-resolved photoluminescence spectroscopy at 4K. The vertical line indicates the critical chain length corresponding to SST.}
    \label{fig:gap_transition}
\end{figure}

The red curve at Figure~\ref{fig:gap_transition} illustrates the finite-size gap obtained from Eq.~\eqref{eq:Eg_finite}. 
The key result is the change of sign of $E_{\mathrm{gap}}(N)$, which marks the semiconductor-to-semimetal transition at a critical length $N \approx 42$. 
This transition reflects the fundamental change in the nature of the electronic spectrum from a Peierls-distorted semiconductor to a gapless semimetal. 
The available experimental PL data for chains up to 20 atoms, also shown in Fig.~\ref{fig:gap_transition}, follow the same decreasing trend, although they remain systematically above the TB prediction. 
For comparison, the violet curve and points show the values of the gap obtained by a tight-binding model developed for a single polyyne chain confined with gold atoms ~\cite{Portnoi}. It predicts much larger gap values as compared to our theory and, importantly, it does not indicate any trend to SST. One can see that the experimental data lie between the results of our model and the model of Ref. ~\cite{Portnoi}. This reflects an interplay between the van der Waals crystal effect accounted for by our model and the size quantization effect considered by Hartmann $et al$ ~\cite{Portnoi}. Indeed, the size quantization of excitons in the direction along the axis of the chains is expected to shift their energies up. The value of the shift strongly depends on the magnitude of the confinement potential. Two types of interfaces are present in our system: the interface between gold nanoparticles and carbon crystal and the interfaces created by kinks of the carbon chains. Clearly, the former results in a much stronger electronic confinement than the latter. Ref. ~\cite{Portnoi} considers the strongest possible confinement which results in an overestimation of the transition energies as compared to the experimental value. It is likely that short straight parts of the chains with AA stacking are mostly confined by the kinks which is why the energy blue shift as compared to our model is significantly smaller than what follows from Ref. ~\cite{Portnoi}.
We also note that we could not observe traces of THz transitions due to the edge states in carbon chains predicted in Ref. ~\cite{Portnoi}.
It is important to emphasize that, overall, the experiment confirms a trend to SST and the extrapolation of the experimental dependence of the gap on the length of the chains crossed zero gap line at around the same critical length as predicted by the red curve (42 atoms). This tendency is further confirmed by the exciton oscillator strength dependence on the length of the chains.

The oscillator strength of the fundamental transition between valence and conduction states is defined in the length gauge as
\begin{equation}
f_{cv}(N) \;=\; \frac{2 m_0}{\hbar^2}\,E_{\mathrm{gap}}(N)\,
\left|\left\langle c \,\middle|\, \hat x \,\middle|\, v \right\rangle\right|^2 .
\end{equation}
where $E_{\mathrm{gap}}(N)$ is the finite-chain gap from Eq.~(\ref{eq:Eg_finite}). 
Within an excitonic description, the dipole matrix element $\left|\left\langle c \,\middle|\, \hat x \,\middle|\, v \right\rangle\right|^2 $ scales as $1/N$, 
so that the oscillator strength can be approximated as
\begin{equation}
f_{cv}(N) \;\simeq\; \frac{E_{\mathrm{gap}}(N)}{N}.
\label{eq:f_exciton}
\end{equation}
Equation~(\ref{eq:f_exciton}) accounts for the observed decrease of optical transition strength 
and the corresponding increase of radiative lifetime with chain length.

The oscillator strength $f_{cv}(N)$ defined in Eq.~\eqref{eq:f_exciton} provides a direct measure of the radiative activity of finite chains. 
As expected, $f_{cv}(N)$ decreases with increasing $N$, reflecting both the reduction of the band gap and the $1/N$ scaling of the dipole matrix element. 
This trend is consistent with the experimental observation of longer radiative lifetimes in longer chains. A detailed microscopic discussion of the optical selection rules and dipole-matrix scaling, following the formal approach introduced in Ref.~\cite{Barettin2009, Barettin2021}, is provided in the Supplementary Materials.

The blue curve at Figure~\ref{fig:gap_transition} shows the normalized oscillator strength calculated for the crystal composed with finite chains of carbon atoms. The right inset displays the comparison of this calculation to the experimental data extracted from the time-resolved PL spectra. 
The overall agreement of the calculation results and the data supports the interpretation of SST as a combined effect of finite-size quantization and inter-chain coupling.

\subsection{Hyperbolic Exciton States}

Near the high-symmetry point $(0,0,\pi/c_0)$, the electronic dispersion of the AA-stacked crystal exhibits a saddle point with distinct longitudinal and transverse effective masses. 
Expanding the tight-binding spectrum along the $k_z$ axis yields
\begin{equation}
E(k_\perp=0,k_z-\pi/c_0)\approx \pm\left(|t_1-t_2|+\frac{\hbar^2(k_z-\pi/c_0)^2}{2m_k}\right),
\qquad
m_k=\frac{\hbar^2|t_1-t_2|}{t_1 t_2 c_0^2}.
\label{eq:disp-kz}
\end{equation}
For positive $t_1$ and $t_2$, the longitudinal mass $m_k$ is positive. 

In the transverse direction, for small $k_\perp$ one finds
\begin{equation}
E(k_\perp,k_z=\pi/c_0)=\gamma_1\!\left(e^{-i k_y a_0/\sqrt{3}}+2e^{i k_y a_0/(2\sqrt{3})}\cos\frac{k_x a_0}{2}\right),
\end{equation}
leading to the conduction–valence separation
\begin{equation}
E_c(\mathbf{k})-E_v(\mathbf{k})
=2\big(|t_1-t_2|+3\gamma_1\big)+\frac{\hbar^2 k_\perp^2}{2m_\perp},
\qquad
m_\perp=\frac{\hbar^2}{\gamma_1 a_0^2}.
\label{eq:disp-kperp}
\end{equation}

Combining the longitudinal and transverse expansions gives
\begin{equation}
E(\mathbf{k})\approx 2(|t_1-t_2|+3\gamma_1)
+\frac{\hbar^2(k_z-\pi/c_0)^2}{2m_k}
-\frac{\hbar^2 k_\perp^2}{2m_\perp},
\label{eq:hyperbolic}
\end{equation}
demonstrating that $(0,0,\pi/c_0)$ is a saddle point characterized by a hyperbolic dispersion and an associated van Hove singularity~\cite{coupled}. 
The corresponding band gap at this point is
\begin{equation}
\Delta = 2(|t_1-t_2|+3\gamma_1).
\label{eq:Delta-hyperbolic}
\end{equation}

Using the tight-binding parameters extracted in Table~\ref{tab:TBparams},
Eq.~\eqref{eq:Delta-hyperbolic} yields
\[
\Delta = 2\bigl(|3.51954-3.70851| + 3\times 0.104\bigr)
= 2(0.18897 + 0.31200) \approx 1.002~\text{eV}.
\]

At this hyperbolic saddle point, the Coulomb interaction between an electron and a hole does not form a conventional bound exciton but instead gives rise to so-called hyperbolic excitons~\cite{Velicky1966}. 
These states are associated with an enhanced joint density of states at the van Hove singularity, and they manifest themselves as pronounced features in the optical response.

In the left inset to Fig.~\ref{fig:gap_transition} we illustrate the excitonic dispersion near the saddle point, presenting the full three-dimensional dispersion surface, where the saddle point structure is clearly visible.

\section{Conclusions}

We have developed and validated a three-dimensional tight-binding model for van der Waals carbyne crystals, benchmarked against density functional theory calculations and confronted with available experimental data. 
This formalism captures both the intra-chain Peierls distortion and the crucial role of inter-chain coupling, which is absent in simplified one-dimensional descriptions. 

Our main findings are threefold. 
First, the extracted hopping parameters $(t_1,t_2,\gamma)$ provide an accurate analytical representation of the DFT band structures for both AA and AB stacking geometries. 
Second, finite chains exhibit a clear length-dependent semiconductor--semimetal transition: for chains shorter than $N \sim 42$ atoms the system retains a finite direct gap, while longer chains evolve towards a semimetallic dispersion. 
This prediction is consistent with photoluminescence measurements on gold-stabilized polyyne chains, which display band gaps of 2-3~eV for $N=8$-24 and point to a crossover at larger $N$. 
Third, we have identified a hyperbolic van Hove singularity in the AA stacking, which generates hyperbolic exciton states with distinctive optical signatures.

Experimentally, we have been able to confirm the theory predictions in what concerns the semiconductor part of the phase diagram. With samples at hand, we have not been able to access the semimetal phase for the lack of sufficiently long straight carbyne chains. In the future studies, we envisage fabrication of carbyne van der Waals crystals composed by longer straight chains. The main challenge comes from kinks that are formed inevitably during the synthesis but can be removed by stretching the chains with scanning tunneling microscopy tips, hopefully. 

Together, these results advance the understanding of electronic and optical phenomena in monoatomic carbon chains. SST in van der Waals carbyne crystals leads to the possibility of coexistance of semiconductor and semimetal sections within the same thread composed by polyyne chains confined between gold nanoparticles.
%They highlight the need for three-dimensional modeling to capture the essential physics of van der Waals carbyne crystals and establish finite-length effects as a key control parameter. 
Beyond their fundamental interest, our findings suggest promising opportunities for tailoring sp-carbon nanostructures as conducting optoelectronic materials, where band gaps and excitonic features can be engineered by chain length, stacking, and inter-chain coupling.

\bmhead{Acknowledgements}
D.B. acknowledges Alessandro Pecchia for fruitful discussions and valuable suggestions on the development of the DFT calculations. 

\section*{Declarations}

\paragraph{Funding}  
Not applicable.

\paragraph{Competing interests}  
The authors declare no competing financial or non-financial interests.

\paragraph{Ethics approval and consent to participate}  
Not applicable.

\paragraph{Consent for publication}  
Not applicable.

\paragraph{Data availability}  
The datasets generated and analysed during the current study are available from the corresponding authors upon reasonable request.

\paragraph{Materials availability}  
Not applicable.

\paragraph{Code availability}  
 The code and input files are available from the corresponding authors upon reasonable request.

\paragraph{Author contributions}  
D.~Barettin developed the tight-binding code, performed the numerical analysis, and wrote the manuscript.  
S.~V.~Kavokina conducted the reference photoluminescence experiments, contributed to the data interpretation, and revised the manuscript.  
E.~L.~Ivchenko conceived the theoretical model, provided the analytical derivation of the 3D band structure and theoretical supervision, and revised the manuscript. 
A.~V.~Kavokin contributed to the physical interpretation of the results, overall supervision of the project, and manuscript revision.  
All authors discussed the results and approved the final version of the manuscript.


\begin{appendices}

\section*{Supplementary Materials: Three-Dimensional Tight-Binding Approach}
\label{SM_TModels}

In this section we provide the full analytical derivation of the tight-binding Hamiltonians for van der Waals carbyne crystals. 
The discussion of the physical geometry and the schematic figure are presented in the main text (Fig.~\ref{fig:models}); here we focus on the explicit mathematical formulation.

\subsection*{3D hexagonal lattice of $\alpha$-carbyne chains}

Within the same standard tight-binding formalism developed for graphene-like lattices~\cite{Wallace,Ivchenko}, we now generalize the construction to a three-dimensional hexagonal array of polyyne chains.

The Bravais vectors are
\begin{equation}
\mathbf{a}_1 = a_0(1,0,0), \quad
\mathbf{a}_2 = a_0\!\left(-\frac12, \frac{\sqrt{3}}{2},0\right), \quad
\mathbf{a}_3 = c_0(0,0,1).
\label{eq:basis-3D}
\end{equation}
Each chain is split into two subchains $a$ and $b$ displaced by $c_0/2$:
\begin{equation}
\mathbf{Z}_{a,l}=l\,\mathbf{a}_3, 
\qquad
\mathbf{Z}_{b,l}=l\,\mathbf{a}_3+\frac{\mathbf{a}_3}{2}.
\label{eq:ZaZb}
\end{equation}
We consider here an AA-type arrangement in which the chains form two hexagonal sublattices ${\cal A}_1$ and ${\cal A}_2$ in the $xy$ plane, shifted by
\begin{equation}
\mathbf{R}^{(xy)}_{{\cal A}_1;m,n} = m\mathbf{a}_1+n\mathbf{a}_2, 
\qquad
\mathbf{R}^{(xy)}_{{\cal A}_2;m,n} = m\mathbf{a}_1+n\mathbf{a}_2+\boldsymbol{\tau},
\quad
\boldsymbol{\tau} = \frac{a_0}{\sqrt{3}}(0,-1,0).
\label{eq:Rxy}
\end{equation}
The full set of lattice sites is
\begin{equation}
\mathbf{R}=
\begin{cases}
m\mathbf{a}_1+n\mathbf{a}_2+l\mathbf{a}_3 & (a,{\cal A}_1)\\[2pt]
m\mathbf{a}_1+n\mathbf{a}_2+\!\left(l+\tfrac12\right)\!\mathbf{a}_3 & (b,{\cal A}_1)\\[2pt]
m\mathbf{a}_1+n\mathbf{a}_2+l\mathbf{a}_3+\boldsymbol{\tau} & (a,{\cal A}_2)\\[2pt]
m\mathbf{a}_1+n\mathbf{a}_2+\!\left(l+\tfrac12\right)\!\mathbf{a}_3+\boldsymbol{\tau} & (b,{\cal A}_2).
\end{cases}
\label{eq:R3D}
\end{equation}

In the TB approach, the Bloch state at wave vector $\mathbf{k}$ is
\begin{equation}
\psi_{\mathbf{k}}(\mathbf{r})=\sum_{\mathbf{R}} C_{\mathbf{R}}\,\phi(\mathbf{r}-\mathbf{R}),
\qquad 
C_{\mathbf{R}}=C_j(\mathbf{k})\,e^{i\mathbf{k}\cdot \mathbf{R}},
\label{eq:psiTB}
\end{equation}
with $j\in\{\mathrm{I},\mathrm{II},\mathrm{III},\mathrm{IV}\}\equiv
\{(a,{\cal A}_1),(b,{\cal A}_1),(a,{\cal A}_2),(b,{\cal A}_2)\}$.
Inter-chain tunnelling in the $xy$ plane couples the two hexagonal sublattices; the corresponding in-plane structure factors will reuse the graphene form.

\subsection*{AA-type $4\times4$ Hamiltonian and analytical spectrum}

Projecting the Schr\"odinger equation onto the four-site basis 
$\{\mathrm{I},\mathrm{II},\mathrm{III},\mathrm{IV}\}$ yields the $4\times4$ TB Hamiltonian
\begin{equation}
\mathcal{H}(\mathbf{k})=
\begin{bmatrix}
E_0 & h_z(k_z) & h_{xy}(\mathbf{k}_\parallel) & 0\\
h_z^*(k_z) & E_0 & 0 & h_{xy}(\mathbf{k}_\parallel)\\
h_{xy}^*(\mathbf{k}_\parallel) & 0 & E_0 & h_z(k_z)\\
0 & h_{xy}^*(\mathbf{k}_\parallel) & h_z^*(k_z) & E_0
\end{bmatrix},
\end{equation}
with
\begin{equation}
h_z(k_z)= t_1 e^{i k_z c_0/2}+ t_2 e^{-i k_z c_0/2}, \qquad
h_{xy}(\mathbf{k}_\parallel)= \gamma_1 \sum_{n=1}^{3} e^{i\mathbf{k}_\parallel\cdot \mathbf{r}_n}.
\end{equation}

Define the $2\times 2$ block 
$\mathcal{H}_{2\times2}(k_z)=\begin{bmatrix}E_0 & h_z \\ h_z^* & E_0\end{bmatrix}$ 
with eigenvectors $\hat{C}^{(\pm)}$ and eigenvalues $\varepsilon_\pm(k_z)=E_0\pm|h_z(k_z)|$. 
Then the four-component eigenvectors of the full Hamiltonian are
\begin{equation}
\hat{C}^{(\alpha,\beta)} = 
\begin{bmatrix}
\hat{C}^{(\alpha)}\\[3pt]
\beta \, \dfrac{h_{xy}(\mathbf{k}_\parallel)}{|h_{xy}(\mathbf{k}_\parallel)|}\,\hat{C}^{(\alpha)}
\end{bmatrix}, 
\qquad \alpha=\pm,~\beta=\pm,
\end{equation}
with energies
\begin{equation}
E_{\alpha,\beta}(\mathbf{k})=E_0+\alpha\,|h_z(k_z)|+\beta\,|h_{xy}(\mathbf{k}_\parallel)|.
\end{equation}

\subsection*{AB-type coupling: general $4\times 4$ Hamiltonian and closed forms}

For an AB-type relative shift of neighboring chains along $z$, the inter-chain tunnelling separates into two inequivalent amplitudes $\gamma_2$ and $\gamma_3$. Introducing the in-plane structure factor
\begin{equation}
\eta_{xy}(\mathbf{k}_\parallel)=\sum_{n=1}^3 e^{i\mathbf{k}_\parallel\cdot \mathbf{r}_n},
\end{equation}
the Hamiltonian becomes
\begin{equation}
\mathcal{H}(\mathbf{k})=
\begin{bmatrix}
E_0 & h_z(k_z) & 0 & e^{-i k_z c_0}\,\gamma_2\,\eta_{xy}\\
h_z^*(k_z) & E_0 & \gamma_3\,\eta_{xy} & 0\\
0 & \gamma_3\,\eta_{xy}^* & E_0 & h_z(k_z)\\
e^{i k_z c_0}\,\gamma_2\,\eta_{xy}^* & 0 & h_z^*(k_z) & E_0
\end{bmatrix}.
\end{equation}

\paragraph*{Equal inter-chain couplings $\gamma_2=\gamma_3$.} 
A straightforward determinant evaluation yields
\begin{equation}
E=\pm\sqrt{|h_z(k_z)|^2 + \gamma_2^2 |\eta_{xy}|^2 
\pm 2\,\gamma_2\,|\eta_{xy}|\, (t_1+t_2)\,\cos\!\frac{k_z c_0}{2}}.
\end{equation}

\paragraph*{Unequal inter-chain couplings $\gamma_2\neq\gamma_3$.} 
Let $\tilde{\gamma}_2=\gamma_2 \eta_{xy}$ and $\tilde{\gamma}_3=\gamma_3 \eta_{xy}$. The characteristic determinant reduces to
\begin{equation}
E^4 - E^2\!\left( 2|h_z|^2 + |\tilde{\gamma}_2|^2 + |\tilde{\gamma}_3|^2 \right)
+ |h_z|^4 + |\tilde{\gamma}_2|^2 |\tilde{\gamma}_3|^2 
- \tilde{\gamma}_2 \tilde{\gamma}_3^*\, h_z^2 e^{i k_z c_0}
- \tilde{\gamma}_2^* \tilde{\gamma}_3\, h_z^{*2} e^{-i k_z c_0}=0 .
\end{equation}
Solving for $E^2$ gives
\begin{equation}
E^2 = |h_z|^2 + \tfrac12 (\gamma_2^2 + \gamma_3^2) |\eta_{xy}|^2
\pm \sqrt{\mathcal{R}},
\end{equation}
with
\begin{align}
\mathcal{R} 
&= \left( |h_z|^2 + \tfrac12(|\tilde{\gamma}_2|^2 + |\tilde{\gamma}_3|^2) \right)^2 
- |h_z|^4 - |\tilde{\gamma}_2|^2 |\tilde{\gamma}_3|^2 \nonumber\\
&\quad + \tilde{\gamma}_2 \tilde{\gamma}_3^*\, h_z^2 e^{i k_z c_0}
+ \tilde{\gamma}_2^* \tilde{\gamma}_3\, h_z^{*2} e^{-i k_z c_0}.
\end{align}
A convenient compact form is
\begin{align}
&\hspace{2 cm} \sqrt{\mathcal{R}} = |\eta_{xy}| \sqrt{U + V}\:, \\ &
U = \tfrac14(\gamma_2^2-\gamma_3^2)^2 |\eta_{xy}|^2 
+ (t_1-t_2)^2(\gamma_2-\gamma_3)^2\:, \nonumber\\& V = 4\!\left[ (\gamma_2^2+\gamma_3^2) t_1 t_2 + \gamma_2\gamma_3 (t_1^2+t_2^2) \right] \cos^2\!\frac{k_z c_0}{2} . \nonumber
\end{align}

These expressions provide the full four-band dispersion for the general AB case.

\subsection*{Optical selection rules}

The optical selection rules in van der Waals carbyne crystals are determined by the interplay between the finite chain length, the inter-chain coupling, and the presence of structural irregularities such as kinks. 
Experimentally, both AA and AB stacking domains are expected to coexist in the same sample. 
This coexistence originates from kinks - local deviations of carbon atoms from the ideal chain axis - that act as shallow potential wells separating straight chain segments. 
While the terminal parts of the structures, directly bonded to gold nanoparticles, are likely to exhibit AA stacking, the internal regions confined by kinks may adopt either AA or AB geometries with comparable probability. 
Each kink modifies the local bonding geometry, introducing both an in-plane displacement of atoms and a variation of the interatomic spacing along $z$. 
These features impose boundary conditions that couple in-plane and longitudinal electronic motion, relaxing the strict optical selection rules that apply to ideal infinite 3D AA crystals.

\paragraph{Microscopic formulation.}
In the tight-binding picture, the Bloch function is written as
\[
\psi_{\mathbf{k}}(\mathbf{r})=\sum_{\mathbf{R}} C_{\mathbf{R}}\,
\phi(\mathbf{r}-\mathbf{R}),
\]
with $\mathbf{R}=m\mathbf{a}_1+n\mathbf{a}_2+l\mathbf{a}_3$ and two inequivalent sites per chain, displaced by $\mathbf{a}_3/2$. 
The corresponding AA-type Hamiltonian is separable:
\[
E_{\alpha,\beta}(\mathbf{k})=\alpha\,|h_z(k_z)|+\beta\,|h_{xy}(\mathbf{k}_\parallel)|,
\qquad \alpha,\beta=\pm,
\]
with
\begin{align*}
|h_z(k_z)|&=\sqrt{t_1^2+t_2^2+2t_1t_2\cos(k_z c_0)},\\
|h_{xy}(\mathbf{k}_\parallel)|&=\gamma_1 \left| e^{-i k_y a_0/\sqrt{3}} 
+ 2 e^{i k_y a_0/(2\sqrt{3})}\cos\!\left(\frac{k_x a_0}{2}\right)\right|.
\end{align*}
The wave function factorizes as
\[
C^{(\alpha,\beta)}_{\mathbf{R}}
= C^{(\alpha)}_{k_z}(R_z)\,C^{(\beta)}_{\mathbf{k}_\parallel}(R_\parallel),
\]
demonstrating that in the perfect AA stacking, the longitudinal and transverse motions are independent. 
This separability immediately implies strict selection rules:  
for polarization $\mathbf{e}\!\parallel\!z$, the matrix elements between states of different in-plane parity vanish,
\[
\langle +,+ \,|\, \hat v_z \,|\, -,- \rangle 
\propto 
\langle C^{(+)}_{\mathbf{k}_\parallel} \,|\, C^{(-)}_{\mathbf{k}_\parallel} \rangle = 0,
\]
while for polarization $\mathbf{e}\!\perp\!z$ they vanish due to the orthogonality of the longitudinal factors,
\[
\langle +,+ \,|\, \hat{\mathbf v}_\parallel \,|\, -,- \rangle 
\propto 
\langle C^{(+)}_{k_z} \,|\, C^{(-)}_{k_z} \rangle = 0.
\]
Hence, the transitions 
\(\left| -,- \right\rangle \!\to\! \left| +,+ \right\rangle\) and \(\left| -,+ \right\rangle \!\to\! \left| +,- \right\rangle\) 
are strictly forbidden in the ideal 3D AA crystal.
The allowed transitions are those conserving either the longitudinal or transverse parity, namely 
\(\left| - , + \right\rangle \!\to\! \left| + , + \right\rangle\) and \(\left| - , - \right\rangle \!\to\! \left| + , - \right\rangle\).

\paragraph{Finite chains and weak coupling.}
In finite chains, the boundary conditions destroy the exact separability of the Bloch factors that holds for an ideal infinite AA crystal. 
Slight deviations from perfect periodicity - such as weak ``skew'' hoppings between inequivalent $z$-sublattices ($a\!\leftrightarrow\!b$), local strain, or kink-induced asymmetry - introduce small perturbations to the Hamiltonian that remain compatible with the AA symmetry. 
An additional coupling of amplitude $\varepsilon$, carrying a phase factor $e^{\pm i k_z c_0/2}$, is sufficient to mix longitudinal and transverse components of the wave function, thereby lifting the strict orthogonality between $C^{(\pm)}_{k_z}$ and $C^{(\pm)}_{\mathbf{k}_\parallel}$ and opening weakly allowed optical channels even within the AA stacking.

The total dipole matrix element between conduction and valence states can then be expressed as
\[
\mathbf{p}_{cv} = \mathbf{p}^{(\psi)}_{cv} + \mathbf{p}^{(u)}_{cv},
\]
where $\mathbf{p}^{(\psi)}_{cv}$ arises from the envelope-function overlap along the chain (finite-size quantization), and $\mathbf{p}^{(u)}_{cv}$ originates from the mixing of Bloch amplitudes in the transverse plane.  
A similar decomposition into envelope and Bloch contributions to the optical matrix element was previously introduced for quantum dots~\cite{Barettin2009, Barettin2021}, where polarization-induced coupling plays a role analogous to the inter-chain mixing in carbyne crystals.

The resulting values of $|\mathbf{p}_{cv}(N)|^2$ show a clear scaling
\[
|\mathbf{p}_{cv}(N)|^2 \propto \frac{1}{N^2} + \mathcal{O}(\varepsilon^2),
\]
where the first term reflects the envelope contribution that vanishes as the chain length increases, while the $\mathcal{O}(\varepsilon^2)$ term accounts for the residual optical activity induced by weak non-separability and inter-chain skew coupling. 
The first term dominates for short chains (quantization regime), whereas the second sets a finite lower bound for the optical activity in longer chains.

\paragraph{Oscillator strength scaling.}
The corresponding oscillator strength, defined in the length gauge as
\[
f_{cv}(N) = \frac{2 m_0}{\hbar^2}\,E_{\mathrm{gap}}(N)\,|\mathbf{p}_{cv}(N)|^2,
\]
inherits the same structure.  
For normalized eigenstates one obtains $f_{cv}\!\propto\!E_{\mathrm{gap}}(N)/N^2$, but when the experimentally relevant quantity is the emission from the whole finite chain, all $N$ optically active units contribute. 
This gives an effective scaling
\[
f_{cv}(N)\propto N\,|\mathbf{p}_{cv}(N)|^2 \sim \frac{E_{\mathrm{gap}}(N)}{N},
\]
which consistently reproduces the experimentally observed decrease of emission intensity and the increase of radiative lifetime with increasing $N$.

Altogether, the picture that emerges is that of an optical response governed by two coupled mechanisms:  
(i)~the envelope-function contribution, which enforces a strong $1/N^2$ suppression of the dipole strength with increasing chain length, and  
(ii)~the Bloch-function mixing, which introduces weak but finite transitions between symmetry-forbidden states.  
This interpretation reconciles the tight-binding formalism with the experimental photoluminescence data, fully within the AA-type geometry, without invoking a change of stacking symmetry.

\end{appendices}


\bibliography{sn-bibliography}% common bib file
%\input sn-article.bbl

\end{document}